\documentclass[submission,copyright]{eptcs}
\providecommand{\event}{ICLP2020} 
\usepackage{breakurl}             
\usepackage{underscore}           
\usepackage{mathptmx}
\usepackage{amsmath,amssymb,amsfonts,mathrsfs}
\usepackage{bm}
\usepackage{graphicx}
\usepackage{booktabs}
\usepackage{algorithm}
\usepackage{algorithmic}
\usepackage{etoolbox}
\usepackage{hyperref,url}
\usepackage{listings}
\usepackage{acronym}
\newacro{SD}{Singly-Defined}
\newacro{AI}{Artificial Intelligence}
\newacro{ANN}{Artificial Neural Network}
\newacro{ASP}{Answer Set Programming}
\newacro{LP}{Logic Programming}
\newacro{COO}{Coordinate}
\newacro{CSR}{Compressed Sparse Row}
\newacro{KB}{Knowledge Base}

\newtheorem{example}{Example}

\newtheorem{definition}{Definition}

\newfont{\Bb}{msbm10}

\title{Enhancing Linear Algebraic Computation of \\ Logic Programs Using Sparse Representation}
\author{Tuan Nguyen Quoc \qquad\qquad  Katsumi Inoue
\institute{National Institute of Informatics, Tokyo, Japan}
\institute{The Graduate University for Advanced Studies (SOKENDAI)}
\email{\{\href{mailto:tuannq@nii.ac.jp}{tuannq}, \href{mailto:inoue@nii.ac.jp}{inoue}\}@nii.ac.jp}
\and
Chiaki Sakama
\institute{Wakayama University, Wakayama, Japan}
\institute{}
\email{\href{mailto:sakama@wakayama-u.ac.jp}{sakama}@wakayama-u.ac.jp}
}
\def\titlerunning{Enhancing Linear Algebraic Computation of Logic Programs Using Sparse Representation}
\def\authorrunning{T. Nguyen, K. Inoue \& C. Sakama}
\begin{document}
\maketitle

\begin{abstract}
  Algebraic characterization of logic programs has received increasing attention in recent years. Researchers attempt to exploit connections between linear algebraic computation and symbolic computation in order to perform logical inference in large scale knowledge bases. This paper proposes further improvement by using sparse matrices to embed logic programs in vector spaces. We show its great power of computation in reaching the fixpoint of the immediate consequence operator from the initial vector. In particular, performance for computing the least models of definite programs is dramatically improved in this way. We also apply the method to the computation of stable models of normal programs, in which the guesses are associated with initial matrices, and verify its effect when there are small numbers of negation. These results show good enhancement in terms of performance for computing consequences of programs and depict the potential power of tensorized logic programs.
\end{abstract}

\section{Introduction}

For decades, \ac{LP} representation has been considered mainly in the form of symbolic logic \cite{kowalski1974logic}, which is useful for declarative problem solving and symbolic reasoning. Logic programming starts gaining more attention recently in order to build explainable learning models, whereas it still has some limitations in terms of computation. In other words, symbolic computation is not an efficient way when we need to combine it with other numerical learning models such as \ac{ANN}. Recently, several studies have been done on embedding logic programs to numerical spaces so that we can exploit great computing resources ranging from multi-threaded CPU to GPU. The linear algebraic approach is a robust way to manipulate logic programs in numerical spaces. Because linear algebra is at the heart of many applications of scientific computation, this approach is promising to develop scalable techniques to process huge relational \ac{KB} \cite{yang2015embedding}, \cite{rocktaschel2014low}. In addition, it enables the ability to use efficient parallel algorithms of numerical linear algebra for computing \ac{LP}.

In \cite{cohen2016tensorlog}, Cohen
described a probabilistic deductive database system in which reasoning is performed by a differentiable process. With this achievement, they can enable novel gradient-based learning algorithms. In \cite{sato2017embedding}, Sato
presented the use of first-order logic in vector spaces for Tarskian semantics. He demonstrates how tensorization realizes the efficient computation of Datalog. In \cite{sato2017linear}, Sato
proposed linear algebraic approach to datalog evaluation. In this work, the least Herbrand model of DB, is computed via adjacency matrices. He also provided theoretical proofs for translating a program into a system of linear matrix equations. This approach achieves $O(N^3)$ time complexity where $N$ is the number of variables in a clause. Continuing to this direction, Sato et al.
developed linear algebraic abduction to abductive inference in Datalog \cite{sato2018abducing}. They did empirical experiments on linear and recursive cases and indicated that this approach can derive base relations.

Using a linear algebraic method, Sakama et al.
\cite{sakama2017linear} define relations between \ac{LP} and multi-dimensional array (tensor) then propose algorithms for computation of \ac{LP} models. The representation is done by defining a series of conversions from logical rules to vectors and then the computation is done by applying matrix multiplication. Later, elimination techniques are applied to reduce the matrix size \cite{nguyen2018computing} and gain impressive performance. In \cite{aspis2018tensor}, a similar idea using 3D-tensor was employed to compute solutions of abductive Horn propositional tasks. In addition, Aspis
built upon previous works on matrix characterization of Horn propositional logic programs to explore how inference from logic programs can be done by linear algebraic algorithms \cite{yaniv2018thesis}. He also proposed a new algorithm for non-monotonic deduction, based on linear algebraic reducts and differentiable deduction. These works prove that the linear algebraic method is promising for logic inference on large scale but has not yet been done much in experiments, to the best of our knowledge.

In this paper, we continue Sakama et al.'s
idea of representing logic programs by tensors \cite{sakama2017linear}. Although the method is well-defined, there are some problems, which limit the performance of the approach and have not been solved. First, the obtained matrix after conversion is sparse but sparsity analysis was not considered. Second, the experiments were limited with small-size logic programs that are not sufficient to prove the robustness of matrix representation. In this research, we further raise the bar of computing performance by using sparse representation for logic programs in order to reach the fixpoint of the immediate consequence operator from the initial vector. We are able to do experiments on larger size logic programs to demonstrate the performance for computing least models of definite programs. Further, we also conduct experiments on computation of stable models of normal programs with a small number of negations.

Accordingly, the rest of this paper is organized as follows: Section \ref{sec:02} reviews and summaries some definitions and computation algorithms for definite and normal programs, Section \ref{sec:03} discusses sparsity problem in tensorized logic programs and proposes a method to represent \ac{LP}, Section \ref{sec:04} demonstrates experimental results with definite and normal programs, Section \ref{sec:05} gives final conclusions and future works.

\section{Preliminaries}
\label{sec:02}

\subsection{Definite programs}

We consider a language $\mathscr{L}$ that contains a finite set of propositional variables. 
Given a logic program $P$, the set of all propositional variables appearing in $P$ is called 
the {\em Herbrand base\/} of $P$ (written $B_P$). 
A {\em definite program\/} is a finite set of {\em rules\/} of the form: 
\begin{equation} \label{h-rule}
  h\leftarrow\; b_1\wedge\cdots\wedge b_m\;\;\;\;  (m\geq 0)
\end{equation}
where $h$ and $b_i$ are propositional variables (atoms) in $\mathscr{L}$.

A rule $r$ is called an {\em OR-rule\/} if $r$ is the form: 
\begin{equation} \label{OR-rule}
  h\leftarrow\; b_1\vee\cdots\vee b_m\;\;\;\;  (m\geq 0)
\end{equation}
where $h$ and $b_i$ are propositional variables in $\mathscr{L}$.

A {\em standardized program\/} is a finite set of rules that are either~(\ref{h-rule}) or~(\ref{OR-rule}). 
Note that the rule~(\ref{OR-rule}) is a shorthand of $m$ rules: $h\leftarrow b_1$, $\ldots$, $h\leftarrow b_m$, so a standardized program is considered a definite program.%

For each rule $r$ of the form~(\ref{h-rule}) or~(\ref{OR-rule}),  
define $head(r)=h$ and $body(r)=\{b_1,\ldots, b_m\}$.%
\footnote{We assume $b_i\neq b_j\; \forall\; i\neq j$.}
A rule $r$ is called a \emph{fact} if $body(r)=\emptyset$. 

\begin{definition}{\textbf{\ac{SD} program}: } \label{sdprogram}
  A definite program $P$ is called a \ac{SD} program if 
  $head(r_1)\neq head(r_2)$ for any two rules $r_1$ and $r_2$ ($r_1\neq r_2$) in $P$. 
\end{definition}

Any definite program can be transformed to a standardized program by introducing new propositional variables. That is, if there are several rules with the same head but different bodies: $h \gets body(r_1)$, $\ldots$, $h \gets body(r_k)$, then we replace all these rules by $h \gets b_1 \vee \ldots \vee b_k$, $b_1 \gets body(r_1)$,... $b_k \gets body(r_k)$.  In this paper, a program means a standardized program unless stated otherwise.

A set $I\subseteq B_P$ is an {\em interpretation\/} of $P$. 
An interpretation $I$ is a {\em model\/} of a standardized program $P$
if $\{b_1,\ldots,b_m\}\subseteq I$ implies $h\in I$ for 
every rule~(\ref{h-rule}) in $P$, and  $\{b_1,\ldots,b_m\}\cap I\neq\emptyset$ implies $h\in I$ for 
every rule~(\ref{OR-rule}) in $P$. 
A model $I$ is the {\em least model\/} of $P$ if $I\subseteq J$ for any model $J$ of $P$. 
A mapping $T_P:\, 2^{B_P} \rightarrow 2^{B_P}$ (called a $T_P$-{\em operator\/}) is defined as: $T_P(I) = \{\, h\,\mid\, h\leftarrow b_1\wedge\cdots\wedge b_m\in P\;\mbox{and}\; \{ b_1,\ldots, b_m \}\subseteq I\,\}\; \cup\; \{\, h\,\mid\, h\leftarrow b_1\vee\cdots\vee b_n\in P\;\mbox{and}\;
  \{ b_1,\ldots, b_n \}\cap I\neq\emptyset\,\}. 
$


The {\em powers\/} of $T_P$ are defined as: 
$T_P^{k+1}(I)=T_P(T_P^k(I))$ $(k\geq 0)$ and $T_P^0(I)=I$. 
Given $I\subseteq B_P$, there is a fixpoint $T_P^{n+1}(I)=T_P^n(I)$ $(n\geq 0)$. 
For a definite program $P$, the fixpoint $T_P^n(\emptyset)$ coincides with the least model of $P$
\cite{van1976semantics}.  

\begin{definition}{\textbf{Matrix representation of standardized programs}{\rm~\cite{sakama2017linear}}: } \label{p-matrix}
  Let $P$ be a standardized program and $B_P=\{ p_1$, $\ldots$, $p_n \}$. 
  Then $P$ is represented by a matrix $M_P\in$ {\Bb R}$^{n\times n}$ such that 
  for each element $a_{ij}$ $(1\leq i,j\leq n)$ in $M_P$, 
\end{definition}
\begin{enumerate}
  \item $a_{ij_k}=\frac{1}{m}\;\; (1\leq k\leq m;\, 1\leq i,j_k\leq n)$ if\,
        $p_i\leftarrow p_{j_1}\wedge\cdots\wedge p_{j_m}$ is in $P$;  
  \item $a_{ij_k}=1\;\; (1\leq k\leq l;\, 1\leq i,j_k\leq n)$ if\,
        $p_i\leftarrow p_{j_1}\vee \cdots \vee p_{j_l}$ is in $P$;
  \item $a_{ii}=1$ if $p_i\leftarrow$ is in $P$;
  \item $a_{ij}=0$, otherwise.
\end{enumerate}
$M_P$ is called a {\em program matrix\/}. 
We write ${\sf row}_i(M_P)=p_i$ and ${\sf col}_j(M_P)=p_j$ $(1\leq i, j\leq n)$.%

To better understand Definition \ref{p-matrix}, let's consider a concrete example.
\begin{example}{\normalfont{Consider the definite program} $P = \left\{ p \gets q \wedge r,\; p \gets s \wedge t,\; r \gets s,\; q \gets t,\; s \gets,\; t \gets \right\}$}. \\ \rm
  $P$ is not an \ac{SD} program because there are two rules $p \gets q \wedge r$ and $p \gets s \wedge t$ having the same head, then $P$ is transformed to the standardized program $P'$ by introducing new atoms $u$ and $v$ as follows: $P' = \{ u \gets q \wedge r,\; v \gets s \wedge t,\; p \gets u \vee v,\; r \gets s,\; q \gets t,\; s \gets,\; t \gets \}$. Then by applying Definition \ref{p-matrix}, we obtain: \\ 
  \[\small
    \bordermatrix{
      ~ & p & q & r & s & t & u & v \cr
      p & 0 & 0 & 0 & 0 & 0 & 1 & 1 \cr
      q & 0 & 0 & 0 & 0 & 1 & 0 & 0 \cr
      r & 0 & 0 & 0 & 1 & 0 & 0 & 0 \cr
      s & 0 & 0 & 0 & 1 & 0 & 0 & 0 \cr
      t & 0 & 0 & 0 & 0 & 1 & 0 & 0 \cr
      u & 0 & 1/2 & 1/2 & 0 & 0 & 0 & 0 \cr
      v & 0 & 0 & 0 & 1/2 & 1/2 & 0 & 0 \cr
    }
  \]
  \label{ex:program}
\end{example}

Sakama et al.
further define representation of interpretation by using \emph{interpretation vector} (Definition \ref{intevector}). This vector is used to store the truth value of all propositions in $P$. The starting point of \emph{interpretation vector} is defined as the \emph{initial vector} (Definition \ref{initvector}).

\begin{definition}{\textbf{Interpretation vector}{\rm~\cite{sakama2017linear}}: } \label{intevector}
  Let $P$ be a program and $B_P=\{ p_1,\ldots,p_n \}$. 
  Then an interpretation $I\subseteq B_P$ is represented by a vector 
  $v=(a_1,\ldots,a_n)^{\sf T}$ where 
  each element $a_i$ $(1\leq i\leq n)$ represents the truth value of the proposition $p_i$ such that 
  $a_i=1$ if $p_i\in I$; otherwise, $a_i=0$. 
  We write ${\sf row}_i(v)=p_i$.
\end{definition}

\begin{definition}{\textbf{Initial vector}: } \label{initvector}
  Let $P$ be a program and $B_P=\{ p_1,\ldots,p_n \}$. 
  Then the {\em initial vector\/} of $P$ is an interpretation vector 
  $v_0=(a_1,\ldots,a_n)^{\sf T}$
  such that $a_i=1$ $(1\leq i\leq n)$
  if ${\sf row}_i(v_0)=p_i$ and a fact $p_i\leftarrow$ is in $P$; otherwise, $a_i=0$. 
\end{definition}

In order to compute the least model in vector space, Sakama et al.
proposed an algorithm which is equivalent to the result of computing least models by the $T_P$-operator. This algorithm is presented in Algorithm \ref{alg:algorithm}.

\begin{definition}{\textbf{$\theta$-thresholding}: } \label{thres}
  Given a value $x$, define $\theta(x) = x'$ where $x' = 1$ if $x \geq 1$; otherwise, $x' = 0$.
  
  Similarly, the $\theta$-{\em thresholding\/} is extended in an element-wise way to vectors and matrices.
  
\end{definition}

\begin{algorithm}[ht]
  \caption{Matrix computation of least model}
  \label{alg:algorithm}
  \textbf{Input}: a \emph{definite program} $P$ and its Herbrand base $B_P = \{p_1, p_2, ..., p_n\}$\\
  \textbf{Output}: a vector $v$ representing the least model
  \begin{algorithmic}[1] 
    \STATE transform $P$ to a \emph{standardized program} $P^{\delta} = Q \cup D$ with $B_{P^{\delta}} = \{p_1, p_2, ..., p_n, p_{n+1}, ..., p_m\}$ where $Q$ is an \ac{SD} program and $D$ is a set of OR-rules.
    \STATE create matrix $M_{P^{\delta}} \in \mathbb{R}^{m \times m}$ representing $P^{\delta}$
    \STATE create initial vector $v_0 = (v_1, v_2, ..., v_m)^T$ of $P^{\delta}$
    \STATE $v = v_0$
    \STATE $u = \theta(M_{P^{\delta}} v)$ \COMMENT{refer to Definition \ref{thres}}
    
    \WHILE{$u \neq v$}
    \STATE $v = u$
    \STATE $u = \theta(M_{P^{\delta}} v)$ \COMMENT{refer to Definition \ref{thres}}
    \ENDWHILE
    \RETURN $v$
  \end{algorithmic}
\end{algorithm}

\subsection{Normal programs}
Normal programs can be transformed to definite programs as introduced in \cite{alferes2000dynamic}. Therefore, we transform normal programs to definite programs before encoding them in matrices.

\begin{definition}{\textbf{Normal program}: } \label{normalprogram}
  A \emph{normal program} is a finite set of normal rules:
  \begin{equation} \label{def:2}
    h \leftarrow b_1 \wedge b_2 \wedge ... \wedge b_l \wedge \neg b_{l+1} \wedge ... \wedge \neg b_m \ (m \geq l \geq 0)
  \end{equation}
  where $h$ and $b_i (1\leq i\leq m)$ are propositional variables (atoms) in ${\cal L}$.
\end{definition}

$P$ is transformed to a definite program by rewriting the above rule to the following form:
\begin{equation} \label{def:normaltransform}
  h \leftarrow b_1 \wedge b_2 \wedge ... \wedge b_l \wedge \overline{b}_{l+1} \wedge ... \wedge \overline{b}_m \ (m \geq l \geq 0)
\end{equation}
where $\overline{b}_i$ is a new proposition associated with $b_i$.

In this part, we denote $P$ as a normal program with an interpretation $I \subseteq B_P$. The \emph{positive form} $P^{+}$ of $P$ is obtained by applying the above transformation. Since a definite program $P^{+}$ is transformed to its standardized program, then we can apply Algorithm \ref{alg:algorithm} to compute the least model. \cite{alferes2000dynamic} proved that if $P$ is a normal program, $I$ is a stable model of $P$ iff $I^{+}$ is the least model of $P^{+} \cup \overline{I}$. We should note that $I^+$ is interpretation of $P^+$ which is a definite program. We can obtain $P^+$ by applying Algorithm \ref{alg:algorithm} to the transformed program $P^+$. Define $\bar{I} = \{\bar{p}\ |\ p \in B_P \setminus I\}$, then $I^+ = I \cup \bar{I}$.

\begin{definition}{\textbf{Matrix representation of normal programs}{\rm~\cite{nguyen2018computing}}: } \label{normal-matrix}
  Let $P$ be a normal program and $B_P=\{ p_1$, $\ldots$, $p_n \}$ and its \emph{positive form} $P^{+}$ with $B_{P^{+}}=\{ p_1,\ldots,p_n, \overline{q}_{n + 1},\ldots, \overline{q}_m\}$.
  
  Then $P^{+}$ is represented by a matrix $M_P\in$ {\Bb R}$^{m\times m}$ such that 
  for each element $a_{ij}$ $(1\leq i,j\leq m)$:
\end{definition}
\begin{enumerate}
  \item $a_{ii} = 1$ for $n + 1 \leq i \leq m$;
  \item $a_{ij} = 0$ for $n + 1 \leq i \leq m$  and $1 \leq j \leq m$ such that $i \neq j$;
  \item Otherwise, $a_{ij}$ ($1 \leq i \leq n$; $1 \leq j \leq m$) is encoded as in Definition \ref{p-matrix}.
\end{enumerate}
$M_P$ is called a {\em program matrix\/}. 
We write ${\sf row}_i(M_P)=p_i$ and ${\sf col}_j(M_P)=p_j$ $(1\leq i, j\leq n)$.%

\begin{example}{\normalfont{Consider a program} $P = \{ p \gets q \wedge s,\; q \gets p \wedge t,\; s \gets \neg t,\; t \gets,\; u \gets v \}$}. \\ \rm
  First, we need to transform $P$ to $P^{+}$ such that $P^{+} = \{ p \gets q \wedge s,\; q \gets p \wedge t,\; s \gets \overline{t},\; t \gets,\; u \gets v \}$.
  Then applying Definition \ref{normal-matrix}, we obtain: \\ 
  \[\small
    \bordermatrix{
      ~ & p & q & s & t & u & v & \overline{t} \cr
      p & 0 & 1/2 & 1/2 & 0 & 0 & 0 & 0 \cr
      q & 1/2 & 0 & 0 & 1/2 & 0 & 0 & 0 \cr
      s & 0 & 0 & 0 & 0 & 0 & 0 & 1 \cr
      t & 0 & 0 & 0 & 1 & 0 & 0 & 0 \cr
      u & 0 & 0 & 0 & 0 & 0 & 1 & 0 \cr
      v & 0 & 0 & 0 & 0 & 0 & 0 & 0 \cr
      \overline{t} & 0 & 0 & 0 & 0 & 0 & 0 & 1 \cr
    }
  \]
\end{example}

Instead of the initial vector in the case of definite programs, the notion of an initial matrix is introduced to encode positive and negative facts in a program. 

\begin{definition}{\textbf{Initial matrix}{\rm~\cite{nguyen2018computing}}: } \label{initmatrix}
  Let $P$ be a normal program and $B_P=\{ p_1,\ldots,p_n \}$ and its positive form $P^{+}$ with $B_{P^{+}}=\{ p_1,\ldots,p_n,$ $ \overline{q}_{n + 1},\ldots, \overline{q}_m\}$. \\
  The initial matrix $M_0 \in \mathbb{R}^{m\times h} (1 \leq h \leq 2^{m -n})$ is defined as follows:
\end{definition}
\begin{enumerate}
  \item each row of $M_0$ corresponds to each element of $B_P$ in a way that $row_i(M_0) = p_i$ for $1 \leq i \leq n$ and $row_i(M_0) = \overline{q}_i$ for $n + 1 \leq i \leq m$;
  \item $a_{ij} = 1$ ($1 \leq i \leq n$, $1 \leq j \leq h$) iff a fact $q_i \gets$ is in $P$; otherwise $a_{ij} = 0$;
        
  \item $a_{ij} = 0$ ($n + 1 \leq i \leq m$, $1 \leq j \leq h$) iff a fact $q_i$ is in $P$; otherwise, there are two possibilities $0$ and $1$ for $a_{ij}$, so it is either $0$ or $1$.
\end{enumerate}

Each column of $M_0$ is a potential stable model in the first stage. We update $M_0$ by applying matrix multiplication with the matrix representation obtained by Definition \ref{normal-matrix} as $M_{k + 1} = \theta(M_P M_k)$. Then, the algorithm for computing the stable models is presented in Algorithm \ref{alg:algorithmnormal}.

\begin{algorithm}[ht]
  \caption{Matrix computation of stable models}
  \label{alg:algorithmnormal}
  \textbf{Input}: a \emph{normal program} $P$ and its Herbrand base $B_P = \{p_1, p_2, ..., p_n\}$\\
  \textbf{Output}: a set of vectors $V$ representing the stable models or $P$
  \begin{algorithmic}[1] 
    \STATE transform $P$ to a \emph{standardized program} $P^{+}$ with $B_{P^{+}}=\{ p_1,\ldots,p_n, \overline{q}_{n + 1},\ldots, \overline{q}_m\}$.
    \STATE create the matrix $M_P \in \mathbb{R}^{m \times m}$ representing $P^{+}$
    \STATE create the initial matrix $M_0 \in \mathbb{R}^{m \times h}$
    \STATE $M = M_0$, $U = \theta(M_P M)$ \COMMENT{refer to Definition \ref{thres}}
    \WHILE{$U \neq M$}
    \STATE $M = U$, $U = \theta(M_P M)$ \COMMENT{refer to Definition \ref{thres}}
    \ENDWHILE
    
    \STATE $V =$ find stable models of $P$ \COMMENT{refer to Algorithm \ref{alg:findstable}}
    \RETURN $V$
  \end{algorithmic}
\end{algorithm}

\begin{algorithm}[ht]
  \caption{Find stable models of $P$}
  \label{alg:findstable}
  \textbf{Input}: program matrix $M$\\
  \textbf{Output}: a set of vectors $V$ representing the stable models of $P$
  \begin{algorithmic}[1] 
    \STATE $V = \emptyset$
    \FOR{$i$ from $1$ to $h$}
    \STATE $v = (a_1,\ldots a_n, a_{n+1}, \ldots, a_m)^T$ ($i^{th}$-column of $M$)
    \FOR{$j$ from $n + 1$ to $m$}
    \STATE $\overline{q}_j = row_j(M)$
    \FOR{$l$ from $1$ to $n$}
    \IF{$row_l(M) = q_j$}
    \STATE \textbf{if} $a_l + a_j \neq 1$ \textbf{then} break;
    \ENDIF
    \ENDFOR
    \STATE \textbf{if} $l \leq n$ \textbf{then} break;
    \ENDFOR
    \STATE \textbf{if} $j \leq m$ \textbf{then} break;
    \STATE \textbf{else} $V = V \cup \{v\}$
    \ENDFOR
    \RETURN $V$
  \end{algorithmic}
\end{algorithm}

This method requires extra steps on transforming and finding stable models of a program. In addition, the initial matrix size grows exponentially by the number of negations $m - n$. Therefore this representation requires a lot of memory and the algorithm performs considerably slower than the method for definite programs if there are many negations appear in the program. But we will later show that this method still has the advantage when there are a small number of negations.

\section{Sparse representation of logic programs}
\label{sec:03}

The idea of representing logic programs in vector spaces could minimize the work with symbolic computation and utilize better computing performance. Besides that, this method copes with the curse of dimension when a matrix representing logic programs becomes very large. Previous works on this topic only consider dense matrices for their implementation and it seems not very impressive in terms of performance even on small datasets \cite{nguyen2018computing}. In order to solve this problem, this paper focuses on analyzing the sparsity of logic programs in vector spaces and proposes improvement using sparse representation for logic programs.

\subsection{Sparsity of logic programs in vector spaces}

A sparse matrix is a matrix in which most of the elements are zero. The level of sparseness is measured by sparsity which equals the number of zero-valued elements divided by the total number of elements \cite{bunch2014sparse}. Because there are a large number of zero elements in sparse matrices, we can save the computation by ignoring these zero values \cite{gustavson1978two}. According to the conversion method of linear algebraic approach, we can calculate the sparsity of a program $P$ \footnote{We only consider the programs in Definition \ref{p-matrix} and Definition \ref{normal-matrix}.}. This calculation is done by counting the number of nonzero-valued elements of each rule in $P$, then let $1$ minus the fraction of the number of nonzero-valued elements and the matrix size.

By definition, the sparsity of a program $P$ is computed by the following equation:
\begin{equation}
  \displaystyle
  sparsity(P) = 1 - \frac{\displaystyle \sum_{r \in P}{|body(r)|}}{n^2}
  \label{equ:sparsity}
\end{equation}\rm
where $n$ is the number of elements in $B_P$ and $|B_r|$ is the length of body of rule $r$.

Accordingly, the representation matrix becomes a high level of sparsity if matrix size becomes larger while the length of body rule is insignificant. In fact, a rule $r$ in a logic program rarely has a body length approx $n$, therefore, $|B_r| \ll n$. In short, we can say that the matrix representation of a logic program according to the linear algebraic approach is highly sparse.

\subsection{Converting logic programs to sparse matrices}

There are several sparse matrix representations. \ac{CSR} is one of the most efficient, robust and widely adopted by many programming libraries among those \cite{bunch2014sparse}. Hence, in this paper, we propose \ac{CSR} for representing logic programs.

In order to understand the \ac{CSR} format, firstly we need to mention the \ac{COO} format which is simple using 2 arrays of coordinates and 1 array of values. The length of these arrays is equal to the number of nonzero elements. The first array stores the row index of each value, and the second array stores the row and column indices of each value, while the third array stores the values in the original matrix. We can imagine that the $i^{th}$ nonzero element in a matrix is represented by a 3-tuple extracted from these 3 arrays at index $i$. Example \ref{ex:coo} illustrates sparse representation in \ac{COO} format for the program $P$ in Example \ref{ex:program}. We should note that in Example \ref{ex:coo}, zero-based indexing is used.

\begin{example}
  \label{ex:coo}
  \ac{COO} representation for $P$ in Example \ref{ex:program}
  \begin{table}[H]
    \centering
    \begin{tabular*}{0.85\textwidth}{@{\extracolsep{\fill}} | l | c  c  c  c  c  c  c  c  c  c |}
      \hline
      Row index & 0 & 0 & 1 & 2 & 3 & 4 & 5 & 5 & 6 & 6 \\
      \hline
      Col index & 5 & 6 & 4 & 3 & 3 & 4 & 1 & 2 & 3 & 4 \\
      \hline
      Value     & 1.0 & 1.0 & 1.0 & 1.0 & 1.0 & 1.0 & 0.5 & 0.5 & 0.5 & 0.5 \\
      \hline
    \end{tabular*}
  \end{table}
\end{example}

Noticeably, in the row index array, it is possible for a value to be repeated consecutively because the nonzero elements may appear in the same row for many times. We may reduce the size of the row index array by considering \ac{CSR} format. In this format, while the column index and the value arrays remain the same, we compress the row index array by storing the index of the row only where nonzero elements appear. That means we do not need to store 2 consecutive $0$ and 2 consecutive $5$ as in Example \ref{ex:coo}. Instead, we store the index of the next row, then finally point the last index to the end of the row (which equals to the number of nonzero elements). Concretely in the row index array, the first element is starting index which is 0. The last element is an extra element to indicate the end of this array which is equal to the number of nonzero elements. We need 2 consecutive values in row index array to extract the nonzero elements in this row. To be specific, we need to interpret $row\_start$ and $row\_end$ of the $i^{th}$ row from the compressed value in $row\_index$ array: $row\_start_i = row\_index[i], row\_end_i = row\_index[i + 1]$.

Example \ref{ex:csr} illustrates this method. For the first row ($i = 0$), we have $row\_start_0 = 0, row\_end_i = 2$, then we extract 2 values $0$ and $1$ for the nonzero element in the first row. These $start$ and $end$ will be used to extract column index and value of nonzero elements. Similarly, the second row ($i = 1$), we have $row\_start_1 = 2, row\_end_1 = 3$ then we have only one nonzero element at index 2. Continue this interpretation until we reach the final row ($i = 6$), we have $row\_start_6 = 8, row\_end_6 = 10$ then we extract last two nonzero elements at index 8 and 9 for the final row.

\begin{example}
  \label{ex:csr}
  \ac{CSR} representation for $P$ in Example \ref{ex:program}
  \begin{table}[H]
    \centering
    \begin{tabular*}{0.85\textwidth}{@{\extracolsep{\fill}} | l | c  c  c  c  c  c  c  c  c  c |}
      \hline
      Row index & 0 & 2 & 3 & 4 & 5 & 6 & 8 & 10 &  &  \\
      \hline
      Col index & 5 & 6 & 4 & 3 & 3 & 4 & 1 & 2 & 3 & 4 \\
      \hline
      Value     & 1.0 & 1.0 & 1.0 & 1.0 & 1.0 & 1.0 & 0.5 & 0.5 & 0.5 & 0.5 \\
      \hline
    \end{tabular*}
  \end{table}
\end{example}

As we can see in Example \ref{ex:csr}, the row index array now has only 8 indices rather than 10 in Example \ref{ex:coo}. We save storing repeatedly indices in the row index array by storing only the position where it starts and ends. This phenomenon does not always encourage the reduction in terms of array size. Imagine the situation where each row has only one nonzero element, then we save nothing with this representation. Actually for a sparse matrix of the size $m \times n$, the \ac{CSR} format saves on memory only when $nnz < (m (n - 1) - 1) / 2$ (where $nnz$ is number of nonzero elements). Fortunately, in case of linear algebraic method, each rule normally has many atoms in the body, therefore the matrix representation has many nonzero elements in a single row. That is why \ac{CSR} format could considerably save memory. In fact, we can save up to $20\%$ of the size of the row index array using \ac{CSR} format. Accordingly, in our method, we propose to implement \ac{CSR} rather than \ac{COO} not only because it saves more memory, but also because it is widely adopted by many programming libraries.

Because a logic program $P$ is highly sparse, applying Algorithm \ref{alg:algorithm} and Algorithm \ref{alg:algorithmnormal} on sparse representation is remarkably faster than the dense matrix. Moreover, sparse representation saves the memory space as well, therefore enabling the ability to deal with large scale \ac{KB}s. We are going to reveal the performance gain by using sparse representation in the next section.

Note that only the matrix representation of the program is sparse, while the initial vector and the initial matrix are not sparse. Thus, in our methods, we will keep the dense format for interpretation matrices.

\section{Experimental results}
\label{sec:04}
In this section, we conduct two experiments on finding the least models of definite programs and computing stable models of normal programs. In order to evaluate the performance of linear algebraic methods, we complete the implementations of Algorithm \ref{alg:algorithm} and Algorithm \ref{alg:algorithmnormal} with $(i)$ $T_P$-operator and $(ii)$ Clasp (Clingo v5.4.1 running with flag \lstinline{--method=clasp}). Our implementations are done with $(iii)$ dense matrices and $(iv)$ sparse matrices. Except Clasp, all implementations are implemented on C++ with CPU x64 as a targeted device (we do not use GPU accelerated code). In terms of matrix representations and operators, we use Eigen 3 library \cite{eigenweb}. The computer running experiments has the following configurations: CPU: Intel Cote i7-4770 (4 cores, 8 threads) @3.4GHz; RAM: 16GB DDR3 @1333MHz; Operating system: Ubuntu 18.04 LTS 64bit.


Focusing on analyzing the performance of sparse representation, we first evaluate our method by conducting experiments on randomized logic programs. We use the same method of \ac{LP} generation conducted in \cite{nguyen2018computing} that the size of logic program defined by the size $n = |B_P|$ of the Herband base $B_P$ and the number of rules $m = |P|$ in $P$. The rules are uniformly generated based on the length (maximum length is $8$) of rule body according to Table \ref{tab:rules}.

\begin{table}[H]
  \caption{Proportion of rules in $P$ based on the number of propositional variables in their bodies.\vspace{0.2cm}}
  \centering
  \begin{tabular*}{1.0\textwidth}{@{\extracolsep{\fill}} |l|rrrrrrrrr|}
    \hline
    Body length & 0 & 1 & 2 & 3 & 4 & 5 & 6 & 7 & 8 \\
    \hline
    Allocated proportion & $< n/3$
    \footnote{This is the proportion of facts in $P$.}
    & $4\%$ & $4\%$ & $10\%$ & $40\%$ & $35\%$ & $4\%$ & $2\%$ & $1\%$ \\
    \hline
  \end{tabular*}
  \label{tab:rules}
\end{table}
\footnotetext{This is the proportion of facts in $P$.}

We further generate denser matrices in order to analyze the efficacy of the sparse method. While keeping the same proportion of facts and rules with body length are 1 and 2, we generate the rest $70\sim80\%$ rules such that their body length is around $5\%$ of the number of propositions. This method leads to the lower sparsity level of generated matrices with approximate 0.95.


Also based on the generation method for definite programs, we generate normal programs by randomly changing $k$ $(4 \leq k \leq 8)$ literals to negations. The important difference from \cite{nguyen2018computing} is we do experiments on much larger $n$ and $m$, because our method, which is implemented on C++, is dramatically more efficient than Nguyen et al.'s
implementation using Maple. The largest size of the logic program in this experiment reaches thousands of propositions and hundreds of thousands of rules. Further, we also compare our method with one of the best \ac{ASP} solvers - Clasp \cite{gebser2016theory} running in the same environment. All methods are conducted 30 times on each \ac{LP} to obtain mean values of execution time. 

In addition, we also conduct a further experiment using non-random problems with definite programs using transitive closure problem. The graph we use is selected from the \href{http://konect.uni-koblenz.de/networks/}{Koblenz network collection} \cite{kunegis2013konect}. This dataset contains binary tuples and we compute transitive closure of them using the following rules: $path(X,Y) \gets edge(X,Y)$ and $path(X,Y) \gets edge(X,Z) \wedge path(Z,Y)$


\subsection{Definite programs}

The final results on definite programs are illustrated in Table \ref{tab:results}. We can see in the results, Dense matrix method is the slowest method and being unable to run with very large programs. Overall, Sparse matrix method is very efficient which is $10\sim15$ faster than Clasp. We should mention that all the codes are executed on single-threaded CPU without using GPU boost or any other parallel computing techniques.


\begin{table}[ht]
  \newcommand{\myheight}{0.35cm}
  \caption{Details of experimental results on definite programs of $T_P$-operator, Clasp and linear algebraic methods (with dense and sparse representation). $n'$ indicates the actual matrix size after transformation.\vspace{0.2cm}}
  \centering
  \begin{tabular}{|c|rrrr|r|r|r|r|}
    \hline
    No. & $n$   & $m$    & $n'$ \footnote{$n'$ is the size of the Herbrand base of a standardized program} & Sparsity & \textbf{$T_P$-operator} & \parbox{1.8cm}{\centering \textbf{Clasp}} & \parbox{1.8cm}{\centering \textbf{Dense matrix}} & \parbox{1.8cm}{\centering \textbf{Sparse matrix}} \\[\myheight]
    \hline
    1   & 1000  & 5000   & 5788                                                                            & 0.99     & 0.0402                  & 0.1680                                    & 2.0559                                           & \textbf{0.0071}                                   \\
    2   & 1000  & 10000  & 10799                                                                           & 0.99     & 0.1226                  & 0.2940                                    & 17.9986                                          & \textbf{0.0127}                                   \\
    3   & 1600  & 24000  & 25198                                                                           & 0.99     & 0.3952                  & 1.8480                                    & 73.3541                                          & \textbf{0.0357}                                   \\
    4   & 1600  & 30000  & 31285                                                                           & 0.99     & 0.4793                  & 2.5360                                    & 116.1158                                         & \textbf{0.0605}                                   \\
    5   & 2000  & 36000  & 37596                                                                           & 0.99     & 0.7511                  & 3.1690                                    & 155.4312                                         & \textbf{0.0692}                                   \\
    6   & 2000  & 40000  & 41936                                                                           & 0.99     & 0.9763                  & 5.1610                                    & 187.6549                                         & \textbf{0.0675}                                   \\
    7   & 10000 & 120000 & 127119                                                                          & 0.99     & 18.5608                 & 9.0720                                    & -                                                & \textbf{0.3798}                                   \\
    8   & 10000 & 160000 & 167504                                                                          & 0.99     & 25.6532                 & 15.7760                                   & -                                                & \textbf{0.4832}                                   \\
    9   & 16000 & 200000 & 211039                                                                          & 0.99     & 57.0223                 & 19.9760                                   & -                                                & \textbf{0.8643}                                   \\
    10  & 16000 & 220000 & 231439                                                                          & 0.99     & 60.4486                 & 24.7860                                   & -                                                & \textbf{0.9429}                                   \\
    11  & 20000 & 280000 & 297293                                                                          & 0.99     & 104.9978                & 30.5730                                   & -                                                & \textbf{0.9048}                                   \\
    12  & 20000 & 320000 & 337056                                                                          & 0.99     & 108.5883                & 34.4030                                   & -                                                & \textbf{1.0614}                                   \\
    \hline
  \end{tabular}
  \label{tab:results}
\end{table}
\footnotetext{$n'$ is the size of the Herbrand base of a standardized program}

\begin{table}[H]
  \newcommand{\myheight}{0.35cm}
  \caption{Details of experimental results on definite programs (with lower sparsity level) of $T_P$-operator, Clasp and linear algebraic methods (with dense and sparse representation). $n'$ indicates the actual matrix size after transformation.\vspace{0.2cm}}
  \centering
  \begin{tabular}{|c|rrrr|r|r|r|r|}
    \hline
    No. & $n$   & $m$    & $n'$   & Sparsity & \textbf{$T_P$-operator} & \parbox{1.8cm}{\centering \textbf{Clasp}} & \parbox{1.8cm}{\centering \textbf{Dense matrix}} & \parbox{1.8cm}{\centering \textbf{Sparse matrix}} \\[\myheight]
    \hline
    1   & 1000  & 5000   & 5876   & 0.95     & 0.1044                  & 0.3970                                    & 2.3102                                           & \textbf{0.0384}                                   \\
    2   & 1000  & 10000  & 10243  & 0.95     & 0.3613                  & 0.9160                                    & 17.5917                                          & \textbf{0.0519}                                   \\
    3   & 1600  & 24000  & 25712  & 0.95     & 0.9478                  & 2.2540                                    & 70.0931                                          & \textbf{0.1634}                                   \\
    4   & 1600  & 30000  & 31430  & 0.95     & 1.1817                  & 3.0130                                    & 120.5195                                         & \textbf{0.3772}                                   \\
    5   & 2000  & 36000  & 36612  & 0.95     & 1.7335                  & 4.7810                                    & 152.9104                                         & \textbf{0.5499}                                   \\
    6   & 2000  & 40000  & 41509  & 0.95     & 2.0378                  & 6.3260                                    & 192.3609                                         & \textbf{0.6284}                                   \\
    7   & 10000 & 120000 & 125692 & 0.95     & 27.8011                 & 10.8930                                   & -                                                & \textbf{1.0816}                                   \\
    8   & 10000 & 160000 & 166741 & 0.95     & 47.2419                 & 18.6050                                   & -                                                & \textbf{2.2907}                                   \\
    9   & 16000 & 200000 & 210526 & 0.95     & 89.5501                 & 21.7110                                   & -                                                & \textbf{3.7931}                                   \\
    10  & 16000 & 220000 & 230178 & 0.95     & 108.1297                & 28.5370                                   & -                                                & \textbf{4.8605}                                   \\
    11  & 20000 & 280000 & 298582 & 0.95     & 144.8006                & 35.0920                                   & -                                                & \textbf{5.3361}                                   \\
    12  & 20000 & 320000 & 335918 & 0.95     & 183.5328                & 42.8420                                   & -                                                & \textbf{5.9182}                                   \\
    \hline
  \end{tabular}
  \label{tab:resultsdense}
\end{table}

The benchmark on denser matrix are presented in Table \ref{tab:resultsdense}. As can be seen in the results, denser matrices require more computation for Sparse matrix method while it does not affect the same scale on other competitors. Despite of that fact, Sparse matrix method still holds the first place in this benchmark. In terms of analyzing the sparseness level of logic programs, we hardly find a program in which the sparsity is less than 0.97. This observation strongly encourages the use of sparse representation for logic programs.

We next show the comparison for computing transitive closure. We assume that a dataset contains $edges$ (tuples of $nodes$), then first perform grounding two rules of defining $path$. The obtained results are demonstrated in Table \ref{tab:resultstransitiveclosure}. In this non-randomized problem, we can see that the matrix representations are very sprase. Therefore, it is no doubt that Sparse matrix method outperforms Dense matrix method. Accordingly, we only highlight the efficiency of sparse representation and omit the dense matrix approach. Surpringly, Sparse matrix method surpasses Clasp once again in this experiment by a large margin.



\begin{table}[H]
  \caption{Details of experimental results on transitive closure problem of $T_P$-operator, Clasp and sparse representation approach. $n'$ indicates the actual matrix size after transformation.\vspace{0.2cm}}
  \centering
  \newcommand{\mywidth}{2.8cm}
  \newcommand{\myheight}{0.35cm}
  \begin{tabular}{|c|rrrr|r|r|r|r|}
    \hline
    \parbox{\mywidth}{Data name       \\($|V|, |E|$)}   & $n$   & $m$    & $n'$   & Sparsity & \textbf{$T_P$-operator} & \parbox{1.8cm}{\centering \textbf{Clasp}} & \parbox{1.8cm}{\centering \textbf{Sparse matrix}} \\[\myheight]
    \hline
    \parbox{\mywidth}{Club membership \\(65, 95)} & 1200  & 14492  & 15600  & 0.99     & 0.8397                  & 0.3370                   & \textbf{0.0255}        \\[\myheight]
    \parbox{\mywidth}{Cattle          \\(28, 217)}         & 1512  & 20629  & 21924  & 0.99     & 0.9541                  & 0.5060                   & \textbf{0.0365}        \\[\myheight]
    \parbox{\mywidth}{Windsurfers     \\(43, 336)}    & 4324  & 99788  & 103776 & 0.99     & 3.6453                  & 3.3690                   & \textbf{0.1824}        \\[\myheight]
    \parbox{\mywidth}{Contiguous USA  \\(49, 107)} & 4704  & 113003 & 117600 & 0.99     & 4.2975                  & 3.8830                   & \textbf{0.1830}        \\[\myheight]
    \parbox{\mywidth}{Dolphins        \\(62, 159)}       & 7564  & 230861 & 238266 & 0.99     & 12.3067                 & 9.3820                   & \textbf{0.4019}        \\[\myheight]
    \parbox{\mywidth}{Train bombing   \\(64, 243)}  & 8064  & 254259 & 262080 & 0.99     & 15.2257                 & 10.6350                  & \textbf{0.4524}        \\[\myheight]
    \parbox{\mywidth}{Highschool      \\(70, 366)}     & 9660  & 333636 & 342930 & 0.99     & 19.9622                 & 15.8010                  & \textbf{0.6618}        \\[\myheight]
    \parbox{\mywidth}{Les Miserables  \\(77, 254)} & 11704 & 445006 & 456456 & 0.99     & 27.7931                 & 21.9560                  & \textbf{0.8300}        \\
    \hline
  \end{tabular}
  \label{tab:resultstransitiveclosure}
\end{table}

As can be witnessed in the results, Dense matrix method is the slowest, even slower than $T_p$-operator, in terms of computation time due to wasting of computation on a huge amount of zero elements. This could be explained by the high level of sparsity of logic programs provided in Table \ref{tab:results}, Table \ref{tab:resultsdense} and Table \ref{tab:resultstransitiveclosure}. Moreover, large dense matrices consume a huge amount of memory therefore the method is unable to run with large scale matrix size. Overall, sparse matrix method is effective in computing the fixpoint of definite programs. On the other hand, the performance would be improved if we use GPU accelerated code and exploit parallel computing power. The results indicates a potential for logical inference using an algebraic method.

\subsection{Normal programs}

In our current method, the number of columns in the initial matrix (Definition \ref{initmatrix}) grows exponentially by the number of negations, we limit the number of negations in this benchmark by $8$ as specified in the experiment setup.



\begin{table}[H]
  \newcommand{\myheight}{0.35cm}
  \caption{Details of experimental results on normal programs of $T_P$-operator, Clasp and linear algebraic methods (with dense and sparse representation). $n'$ indicates the actual matrix size after transformation.\vspace{0.2cm}}
  \centering
  \begin{tabular}{|c|rrrrr|r|r|r|r|}
    \hline
    No. & $n$   & $m$    & $n'$   & $k$ \footnote{The number of negations.} & Sparsity & \textbf{$T_P$-operator} & \parbox{1.6cm}{\centering \textbf{Clasp}} & \parbox{1.6cm}{\centering \textbf{Dense matrix}} & \parbox{1.6cm}{\centering \textbf{Sparse matrix}} \\[\myheight]
    \hline
    1   & 1000  & 5000   & 6379   & 8                                       & 0.99     & 0.0472                  & 0.3070                                    & 3.9560                                           & \textbf{0.0119}                                   \\
    2   & 1000  & 10000  & 12745  & 8                                       & 0.99     & 0.1838                  & 1.0920                                    & 28.1806                                          & \textbf{0.0178}                                   \\
    3   & 1600  & 24000  & 30061  & 8                                       & 0.99     & 0.5525                  & 3.2760                                    & 105.4931                                         & \textbf{0.0559}                                   \\
    4   & 1600  & 30000  & 36402  & 7                                       & 0.99     & 0.6801                  & 4.3050                                    & 168.8044                                         & \textbf{0.0832}                                   \\
    5   & 2000  & 36000  & 42039  & 5                                       & 0.99     & 1.2378                  & 6.7180                                    & 203.2749                                         & \textbf{0.0897}                                   \\
    6   & 2000  & 40000  & 48187  & 8                                       & 0.99     & 1.5437                  & 7.1800                                    & 256.9701                                         & \textbf{0.0991}                                   \\
    7   & 10000 & 120000 & 171967 & 6                                       & 0.99     & 27.3162                 & 7.6820                                    & -                                                & \textbf{0.7124}                                   \\
    8   & 10000 & 160000 & 207432 & 7                                       & 0.99     & 32.5547                 & 24.6990                                   & -                                                & \textbf{0.8424}                                   \\
    9   & 16000 & 200000 & 250194 & 5                                       & 0.99     & 70.3114                 & 30.7180                                   & -                                                & \textbf{1.5603}                                   \\
    10  & 16000 & 220000 & 278190 & 6                                       & 0.99     & 86.5192                 & 35.4050                                   & -                                                & \textbf{1.8314}                                   \\
    11  & 20000 & 280000 & 357001 & 4                                       & 0.99     & 133.7881                & 50.1970                                   & -                                                & \textbf{1.9170}                                   \\
    12  & 20000 & 320000 & 396128 & 4                                       & 0.99     & 150.3377                & 58.6090                                   & -                                                & \textbf{2.1066}                                   \\
    \hline
  \end{tabular}
  \label{tab:resultsnormal}
\end{table}
\footnotetext{The number of negations.}

\begin{table}[H]
  \newcommand{\myheight}{0.35cm}
  \caption{Details of experimental results on normal programs (with lower sparsity level) of $T_P$-operator, Clasp and linear algebraic methods (with dense and sparse representation). $n'$ indicates the actual matrix size after transformation. $h$ indicates the column size of the initial matrix.\vspace{0.2cm}}
  \centering
  \begin{tabular}{|c|rrrrr|r|r|r|r|}
    \hline
    No. & $n$   & $m$    & $n'$   & $k$ & Sparsity & \textbf{$T_P$-operator} & \parbox{1.6cm}{\centering \textbf{Clasp}} & \parbox{1.6cm}{\centering \textbf{Dense matrix}} & \parbox{1.6cm}{\centering \textbf{Sparse matrix}} \\[\myheight]
    \hline  
    1   & 1000  & 5000   & 6385   & 7   & 0.95     & 0.1680                  & 0.3680                                    & 3.7791                                           & \textbf{0.1133}                                   \\
    2   & 1000  & 10000  & 12294  & 8   & 0.95     & 0.2453                  & 1.4940                                    & 30.0642                                          & \textbf{0.1867}                                   \\
    3   & 1600  & 24000  & 33172  & 7   & 0.95     & 0.6819                  & 3.7830                                    & 102.5389                                         & \textbf{0.2219}                                   \\
    4   & 1600  & 30000  & 35091  & 8   & 0.95     & 0.7741                  & 5.9120                                    & 174.5192                                         & \textbf{0.3462}                                   \\
    5   & 2000  & 36000  & 44145  & 8   & 0.95     & 2.3194                  & 7.1020                                    & 197.3004                                         & \textbf{0.4131}                                   \\
    6   & 2000  & 40000  & 49080  & 7   & 0.95     & 3.2665                  & 8.6690                                    & 250.0876                                         & \textbf{0.4895}                                   \\
    7   & 10000 & 120000 & 181550 & 8   & 0.95     & 36.9532                 & 10.4530                                   & -                                                & \textbf{3.2504}                                   \\
    8   & 10000 & 160000 & 203576 & 6   & 0.95     & 54.1106                 & 33.1920                                   & -                                                & \textbf{4.0186}                                   \\
    9   & 16000 & 200000 & 246159 & 4   & 0.95     & 86.3571                 & 48.1860                                   & -                                                & \textbf{7.2193}                                   \\
    10  & 16000 & 220000 & 282734 & 5   & 0.95     & 106.0275                & 56.9150                                   & -                                                & \textbf{8.3059}                                   \\
    11  & 20000 & 280000 & 365190 & 4   & 0.95     & 163.0558                & 78.1790                                   & -                                                & \textbf{9.0177}                                   \\
    12  & 20000 & 320000 & 387094 & 4   & 0.95     & 202.5501                & 84.3270                                   & -                                                & \textbf{11.5203}                                  \\
    \hline
  \end{tabular}
  \label{tab:resultsnormaldense}
\end{table}

First, we perform benchmarks on normal programs which has $0.99$ sparsity level. Table \ref{tab:resultsnormal} illustrates the execution time in detail. As can be witnessed in the results, Sparse matrix method is still faster than Clasp but with a smaller scale it did in definite programs. It is needed to mention that the initial matrix size is remarkably larger due to the limitation of representation. We have to initialize all possible combinations of an atom which appeared with its negation form in the program. There is no doubt that with a larger number of negations, the space complexity of linear algebraic method is exponential. Accordingly, the performance of Sparse matrix method is only better than Clasp with a small fraction of negations.

In the next experiments, we compare different methods on denser matrix. Table \ref{tab:resultsnormaldense} presents the data for this benchmark. Once again, with a limited number of negations, Sparse matrix method holds the winner position.

Noticeably, execution time on normal programs is generally greater than that on definite programs. This is obvious because we have a larger size of initial matrices as well as the need of extra computation on transforming and finding the least models as described in Algorithm \ref{alg:algorithmnormal}. Then the weakness of the linear algebraic method is that we have to deal with all combinations of truth assignments in order to compute the stable model. Accordingly, the column size of the initial matrix exponentially increases by the number of negations. Thus, in the benchmark on randomized programs, we limit the number of negations for all benchmarks so that the matrix can fit in memory. This limitation will become clearer in real problems which have many negations. This is a major problem that we are investigating to do further research.

\section{Conclusion}
\label{sec:05}
In this paper, we analyze the sparsity of matrix representation for \ac{LP} and then propose an improvement for logic programming in vector space using sparse matrix representation. The experimental results on computing the least models of definite programs demonstrate very significant enhancement in terms of computation performance even when compared to Clasp. This improvement remarkably reduced the burden of computation in previous linear algebraic approaches for representing \ac{LP}. Whereas in finding stable models of normal programs, the efficacy of linear algebraic method is limited due to the representation requires a huge amount of memory to store all possible combinations of negated atoms. In spite of the fact, our method is efficient when there are small numbers of negation. However, matrix computation could be more accelerated using GPU. We have tested our implementation in this way, and obtained expected results too.

Sato's linear algebraic method \cite{sato2017linear} is based on a completely different idea to represent logic programs, where each predicate is represented in one matrix and an approximation method is used to compute the extension of a target predicate of a recursive program. We should note that this approximation method is limited in a matrix size 10,000, while our exact method is comfortable with 320,000. Further comparison is a future research topic, yet we could expect that Sato's method can also be enhanced by sparse representation.

The encouraging results open up rooms for improvement and optimization. In an extended version of this paper, we will add more experimental results including a comparison on other benchmark programs for \ac{ASP}. Potential future work is to apply a sampling method to reduce the number of guesses in the initial matrix for normal programs. An algorithm would be to prepare some manageable size of the initial matrix, and if all guesses fail we do some local search and replace column vectors by new assignments then repeat it until a stable model is found. Further research directions on implementing disjunctive \ac{LP} and abductive \ac{LP} should be considered in order to reveal the applicability of tensor-based approaches for \ac{LP}. Additionally, more complex types of the program should be taken in to account to be represented in vector space, for instance 3-valued logic programs and answer set programs with aggregates and constraints.

\bibliographystyle{eptcs}
\bibliography{references.bib}

\end{document}